\documentclass[journal=aamick, manuscript=article]{achemso}
\usepackage{chemformula}
\usepackage[T1]{fontenc}
\usepackage{lineno}
\usepackage{multirow}
\usepackage[none]{hyphenat}
\newcommand{\fig}[1]{Figure~\ref{#1}}
\newcommand{\figp}[2]{Figure~\ref{#1}#2}
\newcommand{\figpp}[2]{Figure~\ref{#1}#2}

\author{Piotr Tatarczak}
\email{Piotr.Tatarczak@fuw.edu.pl}
\author{Tomasz Fąs}
\author{Jan Pawłowski}
\author{Aleksandra Krystyna Dąbrowska}
\author{Jan Suffczyński}
\author{Piotr Wróbel}
\author{Andrzej Wysmołek}
\email{Andrzej.Wysmolek@fuw.edu.pl}
\author{Johannes Binder}
\affiliation[University of Warsaw]
{Faculty of Physics, University of Warsaw, ul. Pasteura 5, 02-093 Warsaw}

\title[Deterministic hBN bubbles as a versatile platform for studies on single-photon emitters]
  {Deterministic hBN bubbles as a versatile platform for studies on single-photon emitters}


\begin{document}

\begin{abstract}

Single-photon emitters (SPEs) in two-dimensional materials are highly promising candidates for quantum technologies. SPEs in hexagonal boron nitride (hBN) have been widely investigated, but mostly in exfoliated or powder samples that require an activation process, making it difficult to compare studies and reproduce results. Here, we address this problem and propose a platform based on large-area metaloraganic vapour phase epitaxy (MOVPE)-grown hBN, which combines reproducibility and scalability with the ability to readily host SPEs without activation. Through the creation of bubbles via electron-beam irradiation, we achieve additional functionalities, including an interference-mediated enhancement of emission by approximately 100-200~\%, dedicated structures that allow the relocation of individual emitters across different systems, and the opportunity to investigate strain-induced effects. Moreover, in contrast to other gas-filled bubbles that deflate at low temperatures, our bubbles remain stable under cryogenic conditions, allowing studies as a function of temperature. To improve the control over the shape and position of bubbles, we demonstrate a~mask-based method that enables deterministic control over bubble formation. The presented hBN bubbles constitute a versatile platform for reproducible studies of hBN-based emitters, providing a reliable insight into their nature and properties.

\end{abstract}

\section{Introduction}

Solid-state single-photon emitters (SPEs) serve as key building blocks for next-generation technologies such as quantum communication, quantum cryptography, and quantum sensing \cite{O'Brien2009_Solid_SPE, Aharonovich2016_Solid_SPE}. SPEs in two-dimensional (2D) materials have gained significant attention in recent years due to their potential for facile integration into photonic structures \cite{vdw_Geim2013, vdw_Meng2023, vdw_Sakib2024} and for strain-induced tuning of the emission \cite{Iff2019_strain, Mendelson2020_strain, Grosso2017_SPE}. Hexagonal Boron Nitride (hBN) plays a crucial role in this field because it hosts SPEs that cover almost the entire visible spectral range, operate at room temperature, and exhibit high brightness and linear polarization \cite{Tran2016_SPE, Tran2016_SPE2, Chejanovsky2016_SPE, Grosso2017_SPE, Koperski2021_SPE, Fournier2021_SPE, Pelliciari2024, Islam2024_SPE, Kumar2024_SPE, Mejia2025_DAP}. To streamline studies on the properties of hBN-based SPEs, it is essential to establish a scalable platform that readily hosts single-photon emitters, enables emission enhancement and strain-tuning of the emission energy, features characteristic markers allowing the same emitters to be quickly located across different setups, and operates from cryogenic to room temperature.

One of the candidates for such a platform are bubbles in 2D materials, which are commonly observed deformations in graphene, transition metal dichalcogenides (TMDCs) and hBN. They are often formed during the transfer of exfoliated flakes \cite {Khestanova2016_graphene_bubbles, Tyurnina2019_TMDCs_bubbles, Lee2022_hBN_bubbles, Lee2022_bubbles_hBN_TMDC_hBN} or can be created by hydrogen plasma treatment and proton irradiation of 2D materials \cite {He2019_BN_bubbles_plasma_hydrogen, Tedeschi2019_ion_bubbles, Blundo2022_ion_bubbles}. Bubbles formed via these methods have been shown to enhance the Raman response of graphene and TMDCs \cite{Huang2018_Raman_graphene_bubbles, Xiao2022_Raman_graphene_bubbles, Jia2019_Raman_TMDCs_bubbles}, as well as the emission from ultraviolet (UV)-emitting color centers in hBN \cite{Lee2022_hBN_bubbles}, through interference effects. The strain in bubbles has been reported to activate SPEs in TMDC \cite{Tonndorf2015_TMDC_SPE_strain, Chakraborty2016_TMDC_SPE_strain,Cianci2023_SPE_TMDCs_bubbles} and there is also evidence that strain can activate emitters in hBN \cite{Proscia2018_BN_bending, Chen2024_BN_bending}, although another study found no direct correlation between strain and emitter density \cite{Li2021_BN_bending}. In addition, bubbles typically exhibit a non-uniform strain distribution \cite{Yue2012_bubbles_equation, Huang2018_Raman_graphene_bubbles, Blundo2022_ion_bubbles}, which may provide a route to tuning emitter energies.

Here, we address these aspects by studying the impact of bubbles on the optical properties of SPEs in hBN and propose bubbles induced by electron-beam (e-beam) irradiation as a prospective platform. The hBN in this study is grown on 2$^{''}$ sapphire substrates by metal-organic vapor phase epitaxy (MOVPE), which offers both scalability and reproducibility \cite{Li2016_MOVPE, Chugh2018_MOVPE, Dabrowska_2021} in contrast to the commonly employed exfoliated or powder-based samples.

\section{Results and discussion}

\subsection{Formation, geometry and mechanical properties of bubbles}

Formation of bubbles in as-grown MOVPE hBN via electron irradiation is possible due to its specific morphology. During post-growth cooling of the hBN/sapphire system from the growth temperature ($\sim$1400 $^{\circ}$C) down to room temperature, differences in thermal expansion of the sapphire substrate and grown hBN lead to the generation of strain at their interface. To relax this strain, at a certain point, a mesh of wrinkles forms \cite{Tatarczak2024_wrinkles, Iwanski2022_wrinkles, Bera2022_wrinkles, Chugh2018_MOVPE}. It allows water to penetrate the hBN/sapphire interface when the sample is exposed to ambient conditions. Electron irradiation of MOVPE-grown hBN induces the radiolysis of intercalated interfacial water, producing molecular hydrogen among other reaction products, which detaches the epitaxial material from the substrate and leads to bubble formation. More details can be found in Ref. \cite{Binder2023_bubbles}. Alternative to e-beam irradiation, UV laser illumination \cite{Tijent2024_hBN_bubbles_UV} has also been reported to create bubbles in MOVPE-grown hBN on sapphire.

\begin{figure}[H]
    \centering
    \includegraphics[width=8.5 cm]{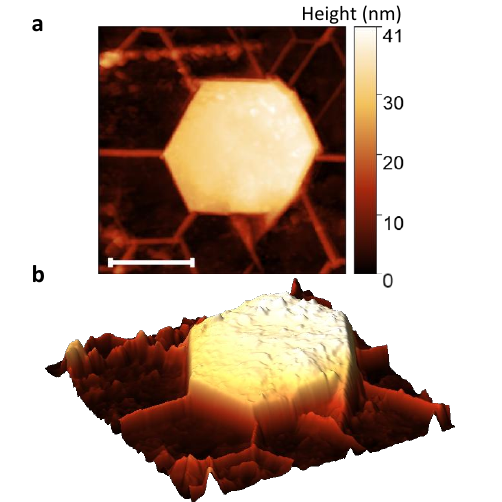}
    \caption{AFM images of a small bubble with the size comparable to a single cell of the wrinkle network: a) top view, b) side view. The scale bar in a) is 500 nm. The total area is 1.5 $\mathrm{\text\textmu m}$ $\times$ 1.5 $\mathrm{\text\textmu m}$.}
    \label{figure1}
\end{figure}

The presence of wrinkles is crucial in the discussed process, so the geometric properties of bubbles are determined by the wrinkle pattern on the hBN surface. In particular, the size of a bubble is not arbitrary, but the smallest possible bubbles are limited to a single one-mesh unit of the wrinkle network, in contrast to other bubbles reported in the literature. An atomic force microscopy (AFM) image of such a one-cell-sized bubble is presented in \fig{figure1}. Enlargement of a bubble requires the detachment of successive hBN mesh units and wrinkles or the merging of existing smaller bubbles, so it is not a continuous process (see Supporting Videos). As shown in \fig{figure1}, the bubble does not exhibit the circular symmetry typical of conventional bubble-like structures, but instead accommodates the wrinkle network. 

The discussed mechanism is completely different from other typical methods of bubble formation in 2D materials. Bubbles are usually formed spontaneously during the transfer of mechanically-exfoliated flakes, when adsorbents are trapped between a thin layer and a new substrate \cite {Khestanova2016_graphene_bubbles, Tyurnina2019_TMDCs_bubbles, Lee2022_hBN_bubbles}. Another approach consists of plasma or proton treatment of exfoliated flakes. In the plasma treatment, ions penetrate the monolayer of the 2D material and induce the formation of molecules that cannot diffuse through the layer and are therefore trapped \cite{He2019_BN_bubbles_plasma_hydrogen, Tedeschi2019_ion_bubbles, Blundo2022_ion_bubbles}. Bubbles formed by both of the above methods can be severely strained due to the high pressure of the trapped gas inside. Depending on bubble size, this pressure can be as high as hundreds of MPa or a few GPa \cite{Khestanova2016_graphene_bubbles}. In such a case, a bubble resembles an inflated balloon, and the layer is stretched. The material that forms the bubble initially covers only a flat circular area, but after bubble formation it is extended over a curved surface. Such a bubble collapses when the temperature falls below the liquefaction point of the enclosed gas, as the internal pressure is the sole factor maintaining its stability \cite{He2019_BN_bubbles_plasma_hydrogen}.

In the case of the bubbles created via e-beam exposure, molecular hydrogen generated in the water radiolysis process is trapped inside the bubble. However, in our case, hydrogen plays a crucial role only during the bubble formation process, and the stiffness of hBN governs the mechanical properties of bubbles afterward. In contrast to conventional bubbles, the presence of additional hBN in the form of wrinkles provides an extra material for the formation of e-beam-induced bubbles. This makes hBN grown at high temperatures particularly prone to detachment from the sapphire, as it allows the layer to relax rather than become strained, as is the case for conventional bubbles. Due to this initial excess of material, bubbles behave like rigid, mechanically stable domes rather than balloons. In particular, they do not collapse at the liquefaction temperature of hydrogen (33.18 K), unlike previously reported hydrogen-filled monolayer hBN bubbles created via plasma treatment \cite{He2019_BN_bubbles_plasma_hydrogen}. Therefore, unlike hBN bubbles created via other methods, the bubbles obtained via electron irradiation serve as an excellent platform to investigate optical properties of bent hBN at cryogenic temperatures. The stability of bubbles below the liquefaction temperature of hydrogen indicates that the contribution of the enclosed gas to the mechanical stability is negligible. On the other hand, since the geometrical shape does not directly depend of the hydrogen pressure inside, the internal pressure cannot be directly assessed.

\begin{figure}[H]
    \centering
    \includegraphics[width=1\textwidth]{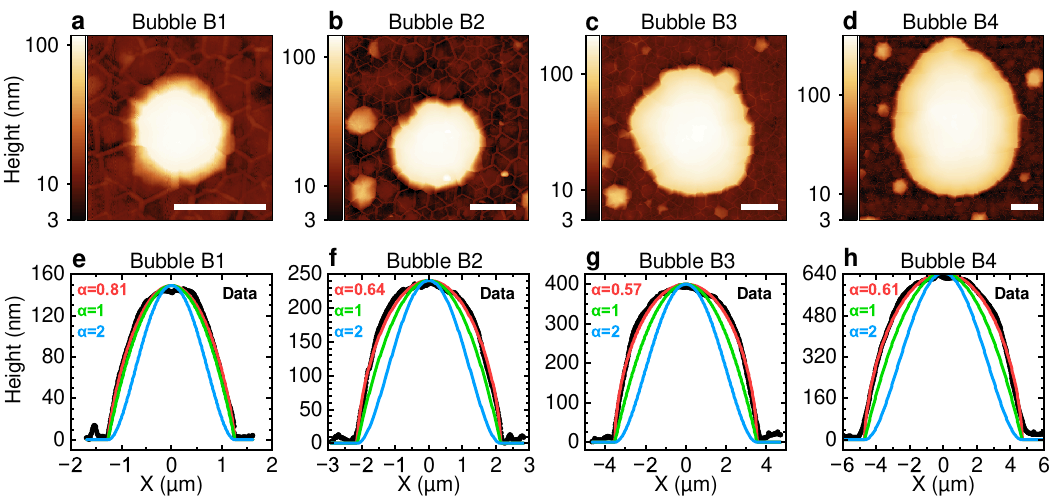}
    \caption{(a-d) AFM images of bubbles B1-B4. Scale bars are 2 $\mathrm{\text\textmu m}$. The logarithmic color scale highlights the presence of wrinkles and smaller bubbles close to the main bubble. (e-h) cross-sectional profiles of bubbles B1-B4 fitted using Eq. \ref{eq1}. Black points - experimental data, green (blue) curves - fits with fixed $\alpha$=1 ($\alpha$=2), red curves - model described in the text.}
    \label{figure2}
\end{figure}

\figpp{figure2}{a-d} show AFM images of four typical bubbles (B1-B4) with respective diameters of $\sim$2.6 $\mathrm{\text\textmu m}$, $\sim$4.3 $\mathrm{\text\textmu m}$, $\sim$7.1 $\mathrm{\text\textmu m}$, and $\sim$9.4 $\mathrm{\text\textmu m}$ fabricated in 18 nm-thick epitaxial hBN by e-beam irradiation. All four bubbles exhibit irregular edges that break the circular symmetry. These edges correspond to the hBN wrinkle network. Smaller bubbles, spanning one or two wrinkle cells, are also visible adjacent to the main bubble. \figpp{figure2}{e-h} present AFM cross-sectional profiles of the discussed bubbles. In the literature, bubble profiles are typically described by: \cite{Khestanova2016_graphene_bubbles, Lee2022_hBN_bubbles, Yue2012_bubbles_equation}

\begin{equation}
    y(x)=H\Bigg(1-\Big(\frac{x}{R}\Big)^{2}\Bigg)^{\alpha},
\label{eq1}
\end{equation}

\noindent
where $y$ is the height at a given position $x$, and $H$ is the maximum height of the bubble of radius $R$. The parameter $\alpha$ describes the mechanical properties and stiffness of the bubble. It should change from $\alpha$=1 for membranes of a negligible thickness to $\alpha$=2 for a finite material of a non-negligible thickness. The green and blue curves in \figpp{figure2}{e-h} show the results of fitting Eq. \ref{eq1} to experimental data with fixed $\alpha$ values (1 and 2, respectively). The model provided with this equation does not accurately describe the shape of the studied bubbles. The profile of the bubbles can be fitted properly only for $\alpha$<1 (red curves) - far from the range predicted by the theory. This discrepancy indicates that the discussed e-beam-induced bubbles exhibit exceptional mechanical properties.

Compared to green and blue theoretical curves, all studied bubbles are slightly broadened at the top. The shape of a bubble is determined by the energetic balance between the strain energy of the hBN, the adhesion to the sapphire substrate, the internal gas energy, and the energy associated with the external pressure \cite{Yue2012_bubbles_equation}. At first glance, it might seem that bubbles are flatter because they were created under vacuum conditions in the electron microscope and, after formation, are influenced by atmospheric pressure. However, one of the key parameters for describing a bubble is its aspect ratio $H/R$. Typically, it remains constant for all bubbles made of the same 2D material, although it may vary slightly between bubbles formed on different substrates. For hBN, this ratio was typically reported as $\sim$0.11 for monolayer or ultrathin bubbles, and slightly lower for thicker bubbles \cite{Khestanova2016_graphene_bubbles, Lee2022_hBN_bubbles, Tijent2024_hBN_bubbles_UV}. $H/R$ ratios of the studied e-beam-induced bubbles fall within the range of 0.11-0.12, in good agreement with literature values. Therefore, we claim that $\alpha$ values smaller than 1, obtained by fitting Eq. \ref{eq1}, are not related to the influence of the external pressure. Instead, we associate such low values with the unusual dome-like behavior of the e-beam-induced bubbles, which originates from their formation mechanism based on the detachment and merging of successive units of the wrinkle network. Having said this, we have to note that the shape of very large bubbles (diameters $\sim$20 $\mathrm{\text\textmu m}$ and larger) can be affected by the ambient pressure and even collapse, but this effect becomes significant only above a critical bubble size, which depends on hBN thickness, morphology, and other factors \cite{Binder2023_bubbles}. For the smaller bubbles investigated here, the intrinsic stiffness of hBN ensures their stability.

\subsection{Impact of bubbles on optical properties of SPEs}

%%%
\begin{figure}[H]
    \centering
    \includegraphics[width=1\textwidth]{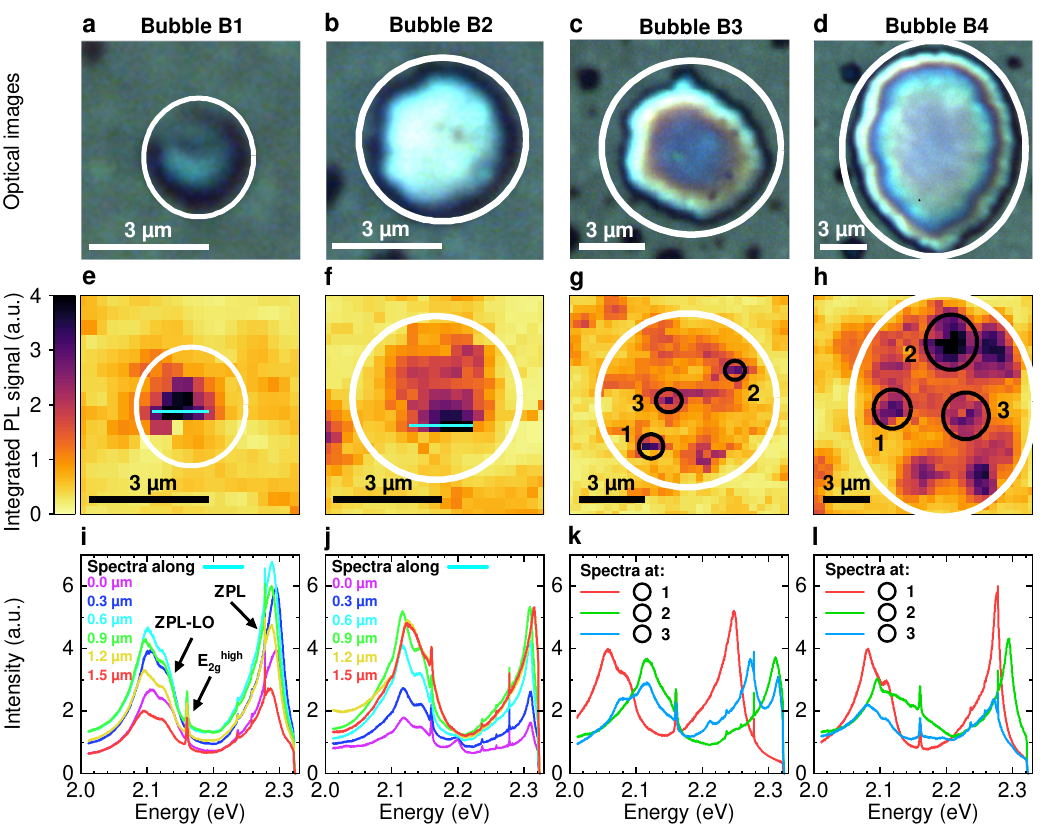}
    \caption{Results of PL mapping measurements conducted on bubbles B1-B4 (each column refers to one bubble). (a-d) optical microscope images of mapped bubbles. (e-h) maps of the total integrated PL intensity in the range 2.0 - 2.33 eV. White circles in (a-h) indicate the bubble positions. Bubbles are rotated compared to \fig{figure2}. Scale bars are 3~$\mathrm{\text\textmu m}$. (i-j)~PL spectra of defects acquired along cyan lines on e) and f) with a 0.3 $\mathrm{\text\textmu m}$ step. (k-l) PL spectra of defects marked by black circles in g) and h). The sharp peaks at $\sim$ 2.28 eV are sapphire Raman modes.}
    \label{figure3}
\end{figure}
%%%

To investigate the impact of bubbles on hBN optical properties, we conducted photoluminescence (PL) measurements. \figpp{figure3}{a-d} present optical microscope images of bubbles B1-B4, and \figpp{figure3}{e-h} show maps of the total integrated intensity of defect-related PL of hBN in the range of 2.0 - 2.33 eV acquired at room temperature. White circles correspond to bubble positions. Generally, the total PL emission is significantly enhanced on bubbles compared to their surroundings. Each bubble features very bright spots, with the corresponding spectra shown in \figpp{figure3}{i-l}. The bright points correspond to SPEs exhibiting an intense zero-phonon line (ZPL) near 2.3 eV. The emission is accompanied by a strong low-energy phonon sideband arising from coupling with acoustic phonons and is followed by a longitudinal optical (LO) phonon replica near 2.1 eV. Small deviations in the emission energy, which are typical for visible point defects in hBN \cite{Tran2016_SPE2, Grosso2017_SPE, Wigger2019_ZPL_deviations}, are observed. Bubbles also naturally serve as markers, making it easy to relocate specific emitters once identified.

\begin{figure}[H]
    \centering
    \includegraphics[width=7.5 cm]{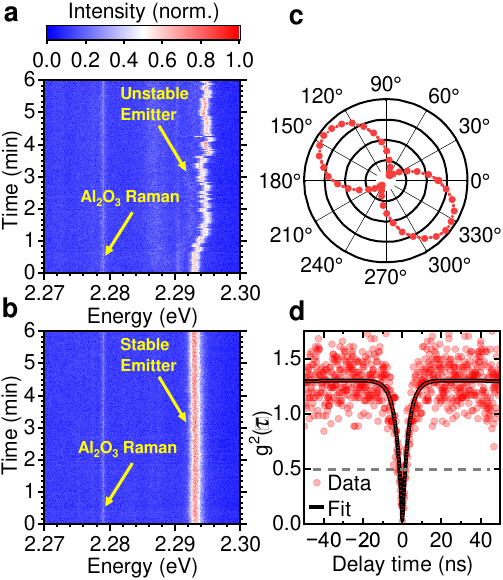}
    \caption{Results of low temperature (1.7 K) PL measurements on single emitters on an hBN bubble. a) and b) time evolution of emission of unstable and stable single-photon emitters, respectively. c) angular dependence of the linearly polarized ZPL emission intensity. d)~second order correlation function curve obtained for the stable single-photon emitter: red circles - experimental points, black curve - fit.}
    \label{figure4}
\end{figure}

Taking advantage of the dome-like behavior of bubbles and their stability at cryogenic temperatures, we performed low-temperature (1.7 K) PL measurements of defect-related emission on bubbles. \figpp{figure4}{a-b} show the evolution of ZPL energy in time at two typical points. The emission of some observed color centers exhibits rapid, discrete spectral jumps (\figp{figure4}{a}). This behavior is similar to reported donor-acceptor-pair-like transitions originating from the competition between recombination processes among neighboring defects \cite{Mejia2025_DAP}. However, for other emitters the ZPL remains stable over a long time (\figp{figure4}{b}). The dependence of the stable emitter emission intensity on polarization angle (shown in \figp{figure4}{c}) reveals a typical in-plane dipole-like emission, as previously reported for SPEs in hBN \cite{Tran2016_SPE, Koperski2021_SPE, Kumar2024_SPE}. Finally, \figp{figure4}{d} shows the results of the single-photon correlation measurement performed on this emitter. The second order correlation function g$^{2}$($\tau$) curve with a clear dip at zero time delay ($\tau$=0) provides evidence for antibunching behavior, confirming that the observed color centers act as true SPEs. The fitted value of g$^{2}$(0)=0.02$\pm$0.05 indicates a high purity of the observed single-photon emission, which is a crucial factor in terms of possible applications. The observed long-time scale bunching at non-zero delays suggests the presence of an additional metastable state involved in the emission process; however, the detailed analysis of the photo-physical properties of the observed SPEs is beyond the scope of this work.

%%%
%%%

%%%
\begin{figure}[H]
    \centering
    \includegraphics[width=7.5 cm]{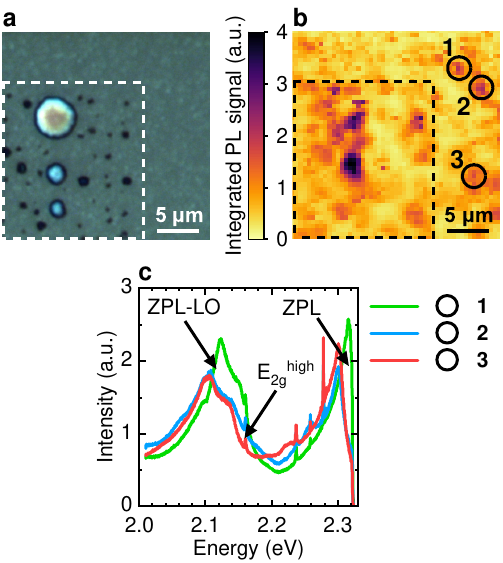}
    \caption{Mapping results for a 25 $\mathrm{\text\textmu m}$ $\times$ 25 $\mathrm{\text\textmu m}$ hBN area. The material in dashed squares was irradiated with an electron beam. a) optical image of the mapped area. b) map of the total integrated PL signal in the range 2.0 eV - 2.33 eV. Scale bars are 5 $\mathrm{\text\textmu m}$. c) spectra acquired in points marked by black circles in b).}
    \label{figure5}
\end{figure}
%%%

An important issue is whether the observed PL enhancement and the presence of intense SPEs are caused by bubble formation or by the e-beam exposure itself. To answer this question, PL mapping measurements  on a partially electron-irradiated hBN area were performed. Results are shown in \fig{figure5}. Only the area marked by a dashed square  was irradiated with an electron beam, while the rest of the mapped area was unaffected. As presented in \figp{figure5}{b}, electron-exposed and unaffected parts of the mapped area do not reveal significant differences in terms of total integrated PL signal, except regions covered by the three largest bubbles. The presence of bubbles does not fully correspond to the exposure pattern (as can be seen in the Supporting Videos); therefore, if electron irradiation causes the formation of new SPEs, we should observe significant PL enhancement throughout the entire area marked by the dashed square. The similarity of these two regions suggests that e-beam exposure itself does not lead to PL enhancement. Moreover, bright spots are observed not only at bubbles but also in the hBN that has not been affected by electrons (points marked by circles in \figp{figure5}{b}). The spectra acquired at such bright points, plotted in \figp{figure5}{c}, closely match those shown in \figpp{figure3}{i-l}, but with 2-3 times lower intensity (intensity scales can be directly compared). This indicates that SPEs are enhanced on bubbles; however, they also occur in our as-grown MOVPE hBN, and no post-growth fabrication is necessary to produce them. This is consistent with prior reports of visible emitters in as-grown hBN synthesized via epitaxy or deposition-based methods \cite{Mendelson2019_SPE_CVD, Stern2019_SPE_hBN_CVD, Koperski2021_SPE, Mendelson2021_hBN_SPE_carbon, Kozawa2023_SPE_hBN_CVD}. Thus, bubble formation facilitates the observation of color centers in hBN only by enhancing their emission. More SPE spectra acquired in flat non-bubbled hBN are shown in Supporting Information in Figure S1. A comparison of spectra presented in \figpp{figure3}{i-l}, \figp{figure5}{c}, and Figure S1 shows that bubble formation can enhance the SPE signal by roughly 100-200\%. This phenomenon is promising for quantum applications in terms of improving the efficiency of SPEs in hBN. 

\begin{figure}[H]
    \centering
    \includegraphics[width=1.0\textwidth]{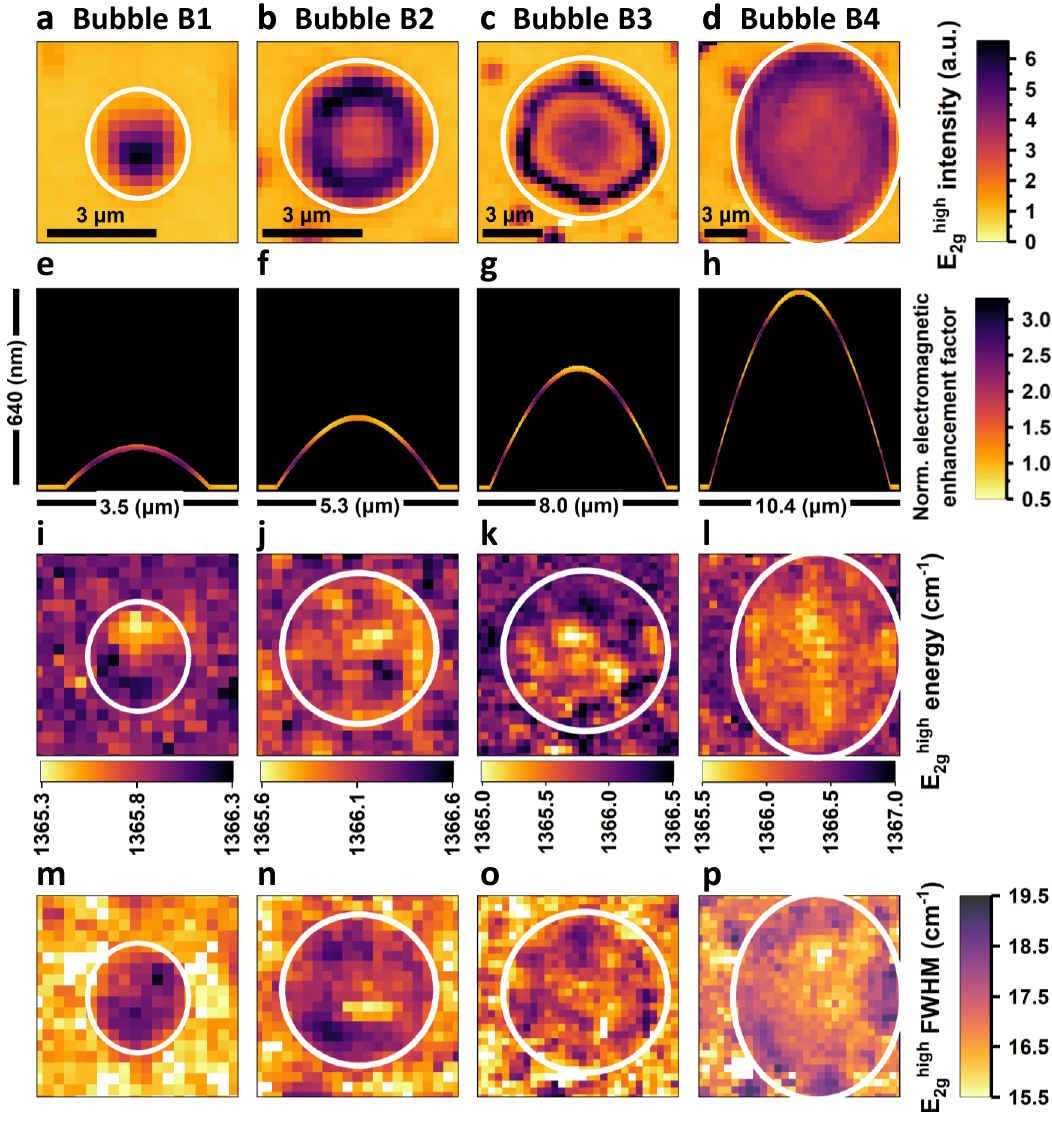}
    \caption{Results of Raman mapping measurements and related optical simulations for bubbles B1-B4. Each column refers to one bubble. (a-d) maps of the intensity of the in-plane E$_{2g}^{high}$ vibrational mode. Scale bars are 3 $\mathrm{\text\textmu m}$. Bubbles are rotated compared to \fig{figure2}. White circles indicate the positions of mapped bubbles. (e-h) Simulated normalized $E_{F}$ within hBN bubbles. (i-l) maps of the phonon energy. (m-p) maps of Raman peak FWHM.}
    \label{figure6}
\end{figure}

Typically, 2D material bubbles exhibit changes in the strain across their surface. The question is whether these strain variations affect SPEs by enhancing their emission or tuning their energy. To assess the strain distribution on the bubbles and investigate its impact on SPEs, we performed Raman mapping measurements. At each point we observed the E$_{2g}^{high}$ in-plane phonon mode of the energy $\sim$ 1370 cm$^{-1}$ typical for sp$^{2}$ hBN \cite{Geick1966_Raman}. Detailed results, shown in \fig{figure6}, reveal pronounced changes in the vibrational properties of hBN induced by bubble formation. Spatially across the bubbles, the Raman signal intensity oscillates, exhibiting up to a tenfold enhancement (\figpp{figure6}{a-d}). Similar oscillations in the Raman intensity, observed in graphene or TMDCs bubbles, were attributed to interference effects \cite{Huang2018_Raman_graphene_bubbles, Xiao2022_Raman_graphene_bubbles, Jia2019_Raman_TMDCs_bubbles}. In particular, for bubble B3, the maxima of the Raman intensity correlate with the Newton's rings observed in the optical image shown in \figp{figure3}{c}. However, such interpretation is less straightforward for the smallest bubble B1, since its height is smaller than half the wavelength corresponding to the Raman signal of the hBN vibrational mode. To understand the influence of bubble geometry on the optical response of hBN in more detail, we performed optical simulations of the spatial variations in electromagnetic enhancement factor $E_{F}=|E/E_{0}|^{4}$, where $E_{0}$ denotes the electric field of the incident wave and $E$ is the local electric field determined by hBN geometry. \figpp{figure6}{e-h} present the distributions of $E_{F}$ in bubble forming hBN layers. The plotted $E_{F}$ is normalized with regard to the flat material next to the bubbles and thus represents the intensity of the Raman signal at a given point relative to that from flat hBN. For all bubbles, we observe $E_{F}$ maxima, which correlate very well with the experimentally observed maxima in Raman intensity shown in \figpp{figure6}{a-d}. We also obtain very good agreement for the smallest bubble, for which the enhancement extends almost across the entire bubble. More detailed simulations, including the $E_{F}$ distribution not only in hBN, but also in the surrounding space, are provided in Figure S2 (Supporting Information).

%%%
%%%
\newpage

\figpp{figure6}{i-l} show the E$_{2g}^{high}$ mode energy across the mapped areas. We observe a general trend that the phonon energy on a bubble is slightly smaller compared to the hBN attached to the substrate. The observed redshift is around 1-2 cm$^{-1}$, in agreement with previously reported changes in hBN Raman on multilayer bubbles \cite{He2019_BN_bubbles_plasma_hydrogen, Lee2022_hBN_bubbles}. Such small shifts indicate that the strain present in bubbles is rather low (of the order of 0.1 \%) \cite{Cai2017_strain, Androulidakis2018_strain}. The formation of hydrogen-irradiation-induced bubbles, exhibiting strain of the order of 1\% and a Raman shift of 50 cm$^{-1}$, was reported in the literature, but this requires detaching only a few or even one layer \cite{Blundo2022_ion_bubbles}, whereas here we deal with a layer of 18 nm thickness (more than 50 layers). Thus, e-beam exposure of our material results in formation of thick, multilayer bubbles revealing smaller strain.

The map of the E$_{2g}^{high}$ mode full width at half maximum (FWHM) is presented in \figpp{figure6}{m-p}. For almost all the bubbles (except B1), the FWHM close to their centers is similar to the values obtained for flat hBN attached to the sapphire. Such a behavior is expected for biaxial strain present at the top of the bubble, which should not lead to a~larger FWHM as the two E$_{2g}^{high}$ modes are not split and remain degenerate. On the other hand, close to the bubble border, the observed FWHM is increased up to $\sim$ 19 cm$^{-1}$ due to the presence of uniaxial strain that breaks the in-plane symmetry, leading to the splitting of phonon modes \cite{Blundo2022_ion_bubbles}.

In contrast to the Raman peak intensity maps, neither the strain distributions nor the FWHM maps show the circular-like symmetry of bubbles. We suppose this is related to the discontinuity in the bubble formation process, the excess of material stored in wrinkles, and the hBN thickness. When two small bubbles merge into a larger one, the curvature of the hBN changes rapidly at the coalescence point, as well as the strain within. However, since the hBN is anchored by wrinkles, the bubble cannot take any shape, but must adapt to the wrinkle pattern. Therefore, relaxation processes are hampered, and some local residual strain may be induced. As a result, the strain distribution does not perfectly reflect the circular symmetry. Moreover, the studied bubbles are surrounded by smaller ones, which can also locally affect the vibrational properties of the material. Intuitively, the smaller a~bubble, the higher the impact of such residual strain. For large bubbles, more wrinkles and wrinkle cells merge, resulting in a more uniform strain distribution. This fact explains why larger bubbles reveal smaller deviations from the expected vibrational properties than the smaller ones (as observed in \figpp{figure6}{i-p}).

To fully take advantage of the bubble-induced SPE enhancement, it is important to understand its true nature and distinguish the influence of strain or interference effects. To determine which phenomenon is responsible for the enhancement, we studied how the total PL signal intensity in the range 2.0 - 2.33 eV correlates with both the E$_{2g}^{high}$ energy and intensity. Since the Raman shift is related to strain, while the intensity of the phonon mode is governed by interference effects, such a comparison may provide an insight into the observed signal enhancement. For each map acquired on bubbles B1-B4 and the one presented in \fig{figure5}, we extracted Pearson correlation coefficients (PCCs) describing linear correlations between the total PL intensity in the range of 2.0 - 2.33 eV and vibrational properties of the E$_{2g}^{high}$ phonon mode. In all cases, only points with the highest total PL intensity related to SPEs were taken into the analysis. These points constitute 25\% of the total points. Spectra with low total intensity were excluded from the analysis to focus only on points where green color centers were present and to avoid the influence of random background fluctuations. The detailed plots are included in Supporting Information (Figure S3), and the obtained PCCs are presented in Table \ref{Table_1}. Our results indicate that the total PL intensity correlates with the Raman peak intensity rather than with its position. In fact, the total PL emission intensity appears, on average, to be uncorrelated with the strain present in bubbles, while it reveals a positive correlation (PCC equal to 0.33) with the Raman peak intensity, which is strongly affected by interference effects. Therefore, we can claim that, in agreement with our simulations and the Raman intensity maps, interference effects are also responsible for the enhancement of the emission of color centers.

\begin{table}[H]
\begin{tabular}{|c|cc|}
\hline
\multirow{2}{*}{Area} & \multicolumn{2}{c|}{Pearson correlation coefficient} \\ \cline{2-3} 
 & \multicolumn{1}{c|}{\begin{tabular}[c]{@{}c@{}}Total PL intensity\\ vs Raman shift\end{tabular}} & \begin{tabular}[c]{@{}c@{}}Total PL intensity\\ vs Raman intensity\end{tabular} \\ \hline
Bubble B1 & \multicolumn{1}{c|}{-0.12} & 0.53 \\ \hline
Bubble B2 & \multicolumn{1}{c|}{0.06} & 0.47 \\ \hline
Bubble B3 & \multicolumn{1}{c|}{0.03} & 0.3 \\ \hline
Bubble B4 & \multicolumn{1}{c|}{0.17} & 0.15 \\ \hline
Partially irradiated area & \multicolumn{1}{c|}{-0.09} & 0.35 \\ \hline
\textbf{Average} & \multicolumn{1}{c|}{\textbf{-0.01}} & \textbf{0.33} \\ \hline
\end{tabular}
\caption{PCCs between the total PL intensity in the range 2.0 - 2.33 eV and vibrational properties.}
\label{Table_1}
\end{table}

\subsection{Formation of bubble matrices}

\begin{figure}[H]
    \centering
    \includegraphics[width=1\textwidth]{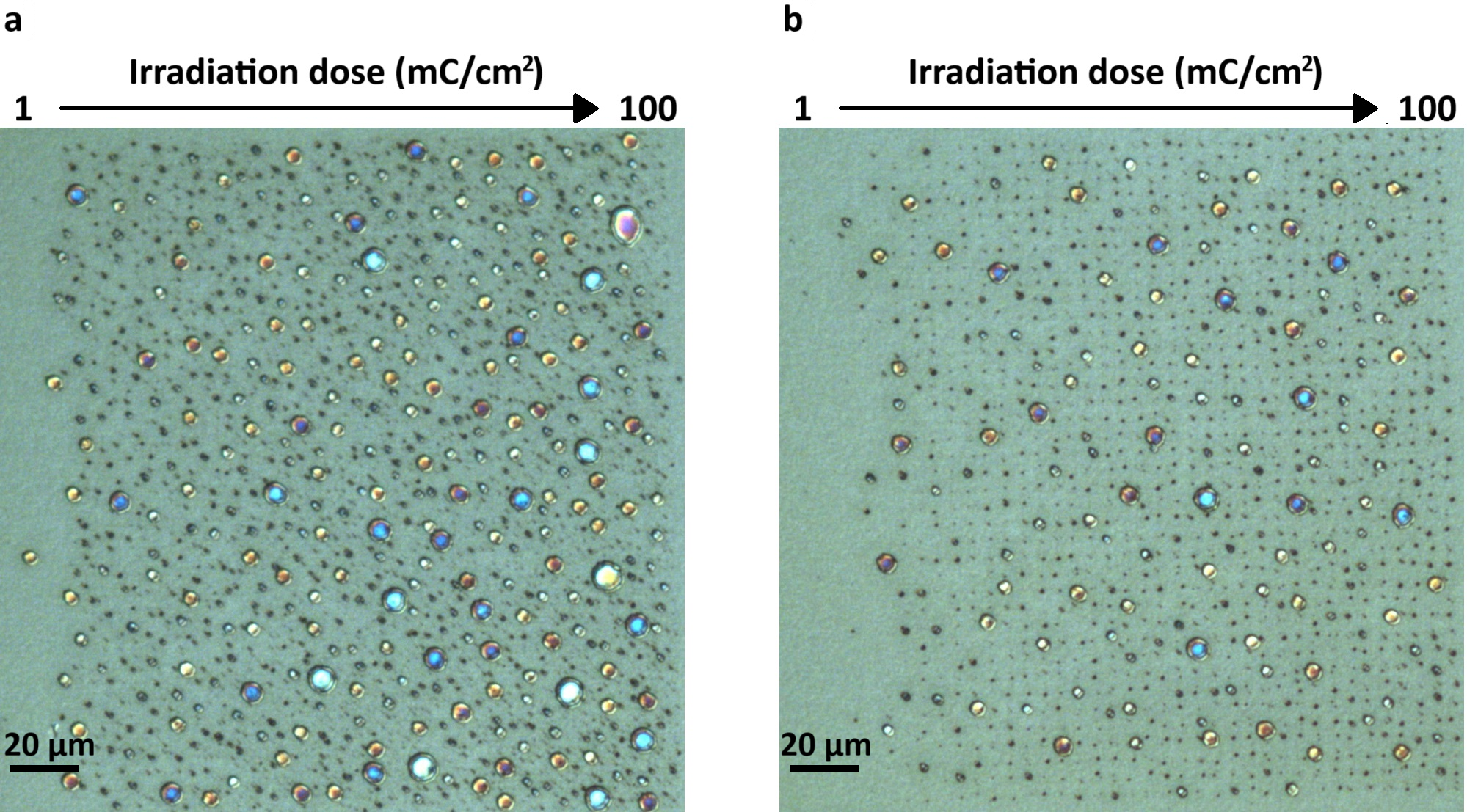}
    \caption{Optical images of 200 $\mathrm{\text\textmu m}$ $\times$ 200 $\mathrm{\text\textmu m}$ matrices 1 and 2 irradiated with an electron beam. Irradiation doses changed from 1 mC/cm$^{2}$ (left sides) to 100 mC/cm$^{2}$ (right sides). Scale bars are 20 $\mathrm{\text\textmu m}$ }
    \label{figure7}
\end{figure}

To leverage the advantages of large-area hBN epilayers over exfoliated flakes or powders, we further address the aspect of scalability of e-beam-induced bubbles. To investigate how irradiation parameters affect the bubble formation, we irradiated matrix patterns using a~dedicated e-beam lithography setup. We exposed circular patterns arranged into two regular 200 $\mathrm{\text\textmu m}$ $\times$ 200 $\mathrm{\text\textmu m}$ matrices. In matrix 1 (matrix 2), the irradiated circles were 1~$\mathrm{\text\textmu m}$ (2 $\mathrm{\text\textmu m}$) in diameter, and the separation between their centers was 2 $\mathrm{\text\textmu m}$ (5 $\mathrm{\text\textmu m}$). The exposure dose applied to each circle in a matrix increased in adjacent columns from 1 mC/cm$^{2}$ to 100~mC/cm$^{2}$ by 1 mC/cm$^{2}$ (2.5 mC/cm$^{2}$) per column. The dose for all circles in a single column was the same. Different irradiation doses were applied by varying the exposure time. \fig{figure7} shows optical images of the hBN layer after exposure. For matrix 1, the created bubbles seem to be distributed randomly over the irradiated area. For matrix 2, the bubble positions are more regular; however, the perfect pre-designed pattern is still not achieved. In the first irradiated region (matrix 1), large bubbles with diameters reaching $\sim$10 $\mathrm{\text\textmu m}$, can be observed, while all bubbles in the second irradiated area are significantly smaller, although applied doses ranged from 1 mC/cm$^{2}$ to 100 mC/cm$^{2}$ for both matrices.

This mechanism of bubble formation, which relies on detaching adjacent units of the wrinkle network and merging small into larger bubbles, can be taken to explain the observed differences between the two matrices. The variation in bubble size arises from different separations between irradiated circles in both matrices. For a 2 $\mathrm{\text\textmu m}$ separation (matrix 1), two or more bubbles can merge into a larger one, causing irregularities in the obtained bubble pattern. For the used doses, a 5 $\mathrm{\text\textmu m}$ step (matrix 2) is too large for the bubbles to merge, so more regular but smaller bubbles are formed. Moreover, variations in the wrinkle pattern on the hBN surface cause bubbles created under identical electron-irradiation conditions to exhibit significant morphological differences. This explains why bubbles within a single column exhibit significant variation, despite identical design parameters. Notably, we observe among different samples that after reaching a specific irradiation time, which varies for different hBN layers, the increase of the bubbles sizes and the formation of new bubbles is stopped. This is in agreement with results on UV-induced bubbles in MOVPE~BN \cite{Tijent2024_hBN_bubbles_UV}. We suppose this critical time is related to the amount of intercalated water. Once all the water in a particular area is radiolyzed, no more hydrogen, which detaches hBN from the sapphire, can be generated. This is another factor that has to be taken into account in the deterministic fabrication of hBN bubbles.

%%%
\begin{figure}[t]
    \centering
    \includegraphics[width=1\textwidth]{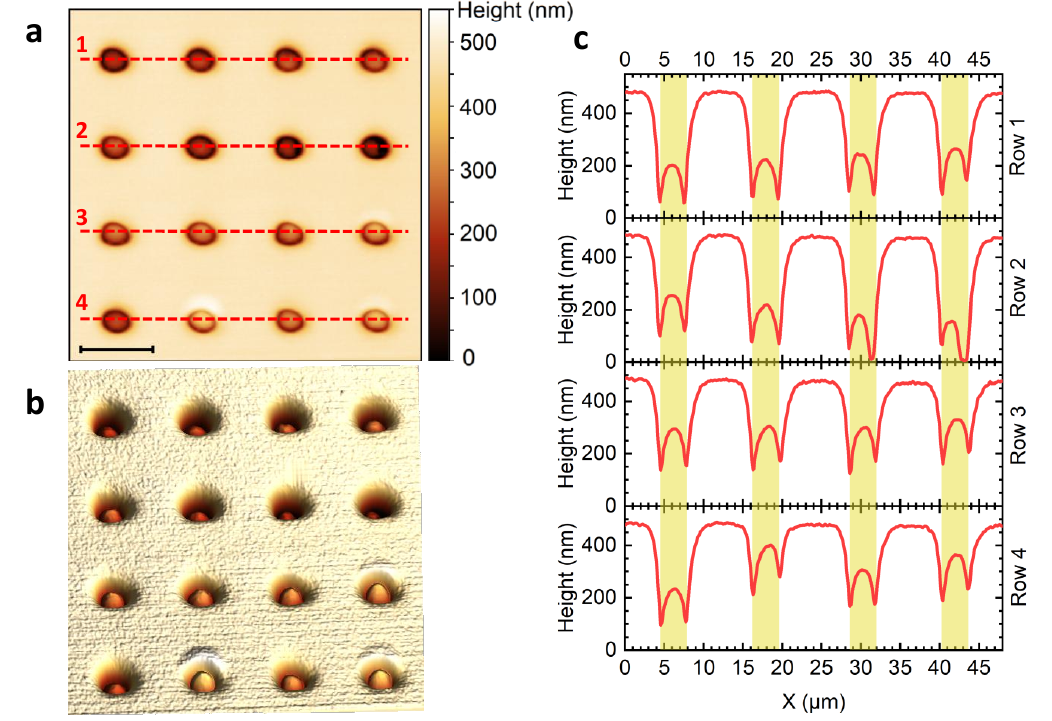}
    \caption{AFM characterization of a representative 4$\times$4 matrix of regular bubbles obtained using a mask-based approach. a) and b) top and isometric views, respectively. The scale bar in a) is 10 $\mathrm{\text\textmu m}$. c) cross-sectional profiles along red dashed lines in a). Bubbles are marked by yellow.}
    \label{figure8}
\end{figure}
%%%

To solve these reproducibility-related issues, we employed a mask-based approach to enforce bubble formation only in selected areas. To this end, we defined a matrix of circular holes into a resist by optical lithography prior to the e-beam irradiation. In the next step, the hBN layer with the mask was irradiated with electrons. \fig{figure8} shows AFM results on the exemplary 4$\times$4 matrix of bubbles designed to be 3 $\mathrm{\text\textmu m}$ in diameter-sized and separated by 12 $\mathrm{\text\textmu m}$. The thickness of the spin-coated resist layer is approximately 440 nm. As expected, the obtained bubbles are lower, so they do not protrude beyond the resist mask. AFM results indicate that our approach allowed us to create separated bubbles at precisely selected areas (resist-free circles). Deviations in the shape of some bubbles are still observed, which obviously require further optimization. Nevertheless, our results constitute a first step toward the deterministic and reproducible fabrication of uniform bubbles that can be used as a versatile platform for studies on SPEs in hBN.

\section{Conclusions}

SPEs in hBN are highly promising candidates for future applications in quantum technologies. Despite intensive scientific efforts, the precise nature of many defect bands remains unidentified. This identification is hindered by studies performed on widely differing materials and activation protocols. In this work, we address this issue and demonstrate a platform that readily hosts SPEs and enables reproducible results through the use of epitaxial MOVPE growth. To add further functionalities, we employed the formation of electron-beam-induced bubbles. This bubble formation is enabled by a wrinkle mesh, which is typical of high-temperature growth of hBN on sapphire. Water that penetrates the hBN/sapphire interface is dissociated by an electron-beam-induced radiolysis process, resulting in the generation of molecular hydrogen, which leads to detachment of the hBN layer. We showed that this formation mechanism gives rise to unusual mechanical properties, as it progresses one wrinkle mesh unit at a time. In contrast to other gas-filled bubbles, the presented bubbles remain stable down to cryogenic temperatures, thereby enabling studies over a wide temperature range. The bubble pattern can also be used to easily relocate a~given emitter across different setups. Using Raman spectroscopy, supported by theoretical simulations, we showed that the shape of the bubbles leads to interference effects that enhance the emission. For the SPE at around 2 eV, we find an emission enhancement of about 100–200~\% on the bubbles, which we show to be related to interference effects rather than strain by analyzing the correlation with the Raman peak intensity and the FWHM. To make full use of the large-area hBN grown by MOVPE, we irradiated matrix patterns of bubbles. This approach, however, reveals difficulties in the deterministic placement of bubbles due to the wrinkle-dominated formation process. To address this problem, we demonstrated a~mask-based irradiation process that allows the generation of bubbles at desired positions.  The results obtained show that bubbles on MOVPE-grown hBN provide a highly versatile platform to study SPEs in hBN in a reproducible and efficient way. The presented approach allows for streamlined research efforts and enables extensive studies needed to unveil the nature of SPEs in hBN.

\section{Experimental section}

\subsection {hBN growth}
The hBN layers were grown on a 2$^{''}$ sapphire c-plane substrate with 0.3$^{\circ}$ offcut by MOVPE using an Aixtron CCS 3$\times$2” system equipped with an ARGUS Thermal Mapping System. The boron and nitrogen precursors were triethylboron (TEB) and ammonia, respectively. Hydrogen was used as a carrier gas, and the growth temperature was 1400 $^{\circ}$C. The sample was grown by a two-stage epitaxy growth process as described in Ref. \cite{Dabrowska_2021}. 

\subsection{Bubble formation}
Bubbles were formed via an e-beam exposure using an FEI Helios 600 Dual Beam system equipped with a RaithElphy electron lithography setup. During the experiment, the beam current was typically 1 nA and the energy of the incident electrons was 5 keV. To create regular bubble matrices, we covered hBN with $\sim$400 nm thick ma-P 1205 photoresist, and then, using UV lithography, we created regular matrices of equally separated holes of the same size. Exposition was performed with a POLOS $\mathrm{\text\textmu}$Printer maskless lithography system.

\subsection{AFM measurements}
AFM experiments were performed using a Dimension Icon microscope equipped with a~NanoScope 6 controller (Bruker Corporation, Billerica, MA, USA). Topographical images were acquired in PeakForce Tapping mode using ScanAsyst-Air probes (Bruker, nominal spring constant: 0.4 N/m, resonance frequency: 70 kHz). Probes were calibrated using the thermal tuning method prior to measurements.

\subsection{Optical measurements}
Micro-Raman and room temperature photoluminescence measurements were performed using a Renishaw inVia setup. The excitation source was a 532 nm continuous-wave (CW) laser. A 100$\times$ objective (NA=0.9) provided a spot size of about 0.5 $\mathrm{\text\textmu m}$. Mapping measurements were acquired using an automated \textit{xyz} stage with a resolution of 100 nm.

During low-temperature PL and the second-order correlation measurements, the sample was mounted inside an Oxford Instruments Spectromag cryostat, which provided a stable low-temperature environment, cooling the sample down to 1.7 K. Optical excitation was achieved using a CW laser operating at a wavelength of 532 nm. The excitation laser beam was focused onto the sample using a lens integrated with a piezoelectric nanopositioning system (Attocube), providing an excitation spot with a diameter of approximately 1.5~$\mathrm{\text\textmu m}$. The PL signal passed through a polarization-resolved detection consisting of a~rotatable half-wave plate followed by a linear polarizer, enabling polarization-dependent measurements. The second-order correlation measurements were performed using a Hanbury Brown and Twiss configuration. Split beams were directed into two high-resolution spectrometers, each equipped with a single-photon counting module (PerkinElmer AQR-14). Through those spectrometers, we select and direct a particular spectral line towards the photodiodes. The time-correlated single-photon counting technique was employed using a HydraHarp 400 time-tagging module, which recorded photon arrival times from both detectors. This allowed for the recording of second-order correlation functions, g$^{2}$($\tau$), with an overall temporal resolution of about 400 ps.

\subsection{Optical simulations}
3D finite-difference time-domain (FDTD) simulations were carried out using Ansys Lumerical FDTD \cite{Lumerical2024}. The computational domain was discretized with a non-uniform Cartesian mesh, with the additional mesh in the regions containing hBN bubbles refined to 1.4 nm to ensure numerical accuracy. Perfectly matched layers (PML) were applied at all simulation boundaries to suppress spurious reflections. The structures were illuminated with a linearly polarized plane-wave source, generated using the total-field scattered-field (TFSF) approach at 573.7 nm, corresponding to the E$_{2g}^{high}$ vibrational mode.

The shape of each bubble was modeled as a custom surface defined by experimentally extracted functions. To accurately reproduce the optical response, the material dispersion of hBN was included using experimentally tabulated refractive indices obtained from spectroscopic ellipsometry (J.A. Woollam RC2 system) measurements of hBN samples used in the preparation of the bubble structures. The experimental spectral data points were subsequently interpolated using Lumerical’s built-in multi-coefficient model. The inner side of each bubble was filled with air.

The sapphire substrate was modeled as a half-infinite medium extending beyond the PML boundaries, corresponding to the single-side polished wafer used in the experiment. The dispersion data for sapphire were taken from Palik’s handbook \cite{Palik1998}.

\begin{acknowledgement}

This work was supported by the Polish National Science Centre under decisions 2020/39/D/ST7/02811 and 2022/45/N/ST7/03355.

\end{acknowledgement}

\bibliography{biblio}

\newpage

\begin{center}
    \textbf{TOC graphic}
    \begin{figure}[H]
    \centering
    \includegraphics[width=1\textwidth]{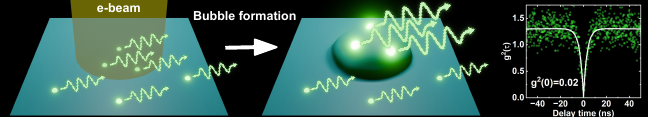}
\end{figure}
\end{center}

Single-photon emitters (SPEs) in hBN are promising for quantum technologies; however, in exfoliated samples their activation is required, limiting reproducibility of previous studies. This work introduces a large-area MOVPE-grown hBN platform that hosts SPEs without prior activation. Electron-beam-created bubbles enhance emission, enable relocation of individual emitters, remain stable at cryogenic temperatures, and can be deterministically positioned, thereby enabling reproducible studies.

\end{document}